\documentclass[prl,twocolumn,amsmath,amssymb,superscriptaddress,floatfix]{revtex4-1}

\usepackage{graphicx}
\usepackage{dcolumn}
\usepackage{bm}
\usepackage[sort&compress]{natbib}
\usepackage{xspace}
\usepackage{amsmath}
\usepackage{epstopdf}
\usepackage{amssymb}
\usepackage{epsfig}

\begin{document}

\title{Quantum Frequency Conversion of a Quantum Dot Single-Photon Source on a Nanophotonic Chip}

\author{Anshuman Singh} \email{anshuman.singh@nist.gov}
\affiliation{National Institute of Standards and Technology, Gaithersburg, MD 20899, USA}
\affiliation{Maryland NanoCenter, University of Maryland, College Park, MD 20742, USA}
\author{Qing Li}
\affiliation{National Institute of Standards and Technology, Gaithersburg, MD 20899, USA}
\affiliation{Maryland NanoCenter, University of Maryland, College Park, MD 20742, USA}
\author{Shunfa Liu}
\affiliation{State Key Laboratory of Optoelectronic Materials and Technologies, School of Electronics and Information Technology, School of Physics, Sun Yat-sen University, Guangzhou, China}
\author{Ying Yu}
\affiliation{State Key Laboratory of Optoelectronic Materials and Technologies, School of Electronics and Information Technology, School of Physics, Sun Yat-sen University, Guangzhou, China}
\author{Xiyuan Lu}
\affiliation{National Institute of Standards and Technology, Gaithersburg, MD 20899, USA}
\affiliation{Maryland NanoCenter, University of Maryland, College Park, MD 20742, USA}
\author{Christian Schneider}
\affiliation{Technische Physik, Universit{\"a}t W{\"u}rzburg, D-97074 W{\"u}rzburg, Germany}
\author{Sven H{\"o}fling}
\affiliation{Technische Physik, Universit{\"a}t W{\"u}rzburg, D-97074 W{\"u}rzburg, Germany}
\affiliation{SUPA, School of Physics and Astronomy, University of St Andrews, St Andrews, United Kingdom}
\author{John Lawall}
\affiliation{National Institute of Standards and Technology, Gaithersburg, MD 20899, USA}
\author{Varun Verma}
\affiliation{National Institute of Standards and Technology, Boulder, CO 80305, USA}
\author{Richard Mirin}
\affiliation{National Institute of Standards and Technology, Boulder, CO 80305, USA}
\author{Sae Woo Nam}
\affiliation{National Institute of Standards and Technology, Boulder, CO 80305, USA}
\author{Jin Liu} \email{liujin23@sysu.edu.cn}
\affiliation{State Key Laboratory of Optoelectronic Materials and Technologies, School of Electronics and Information Technology, School of Physics, Sun Yat-sen University, Guangzhou, China}
\author{Kartik Srinivasan} \email{kartik.srinivasan@nist.gov}
\affiliation{National Institute of Standards and Technology, Gaithersburg, MD 20899, USA}
\affiliation{Joint Quantum Institute, NIST/University of Maryland, University of Maryland, College Park, MD 20742, USA}

\date{\today}

\begin{abstract}
Single self-assembled InAs/GaAs quantum dots are promising bright sources of indistinguishable photons for quantum information science.  However, their distribution in emission wavelength, due to inhomogeneous broadening inherent to their growth, has limited the ability to create multiple identical sources. Quantum frequency conversion can overcome this issue, particularly if implemented using scalable chip-integrated technologies. Here, we report the first demonstration of quantum frequency conversion of a quantum dot single-photon source on a silicon nanophotonic chip.  Single photons from a quantum dot in a micropillar cavity are shifted in wavelength with an on-chip conversion efficiency $\approx~12~\%$, limited by the linewidth of the quantum dot photons. The intensity autocorrelation function $g^{(2)}(\tau)$ for the frequency-converted light is antibunched with $g^{(2)}(0)~=~0.290~\pm~0.030$, compared to the before-conversion value $g^{(2)}(0) =0.080~\pm~0.003$. We demonstrate the suitability of our frequency conversion interface as a resource for quantum dot sources by characterizing its effectiveness across a wide span of input wavelengths (840~nm to 980~nm), and its ability to achieve tunable wavelength shifts difficult to obtain by other approaches.
\end{abstract}
\maketitle

\section{Introduction}

Single photons are fundamental constituents of many quantum technologies~\cite{Kimble2008,Ladd2010,Northup2014,Kurizki2015,Koenderink2015}. Self-assembled InAs/GaAs quantum dots (QDs)~\cite{michler_single_2009}, in particular, have been steadily developed as single-photon sources~\cite{Aharonovich2016,Senellart2017}, to the point that they can now out-perform other sources in simultaneously achieving high brightness, single-photon purity, and indistinguishability~\cite{Somaschi2016,ding_-demand_2016,unsleber_highly_2016,he_deterministic_2017}. As a result, they are relevant to applications that rely on quantum interference of single photons, including linear optics quantum computing~\cite{Kok2007} and more specialized simulations such as Boson sampling~\cite{Aaronson2010,Gard2014}.  Recent Boson sampling experiments using a single QD single-photon source de-multiplexed into a waveguide interferometer network have shown promising potential to scale up the computational complexity that can be addressed in such experiments~\cite{Loredo2017,He2017,Wang2017}. Further progress, not just on Boson sampling but also in other areas such as the construction of multi-photon entangled states, would be greatly aided by increasing the available photon flux through the ability to create multiple identical QD single-photon sources.  However, the inhomogeneous broadening characteristic of self-assembled InAs/GaAs quantum dots~\cite{michler_single_2009} limits the extent to which any two quantum dots can be expected to have the same emission wavelength.
	
\begin{figure*}[t]
		\includegraphics[width=0.75\linewidth]{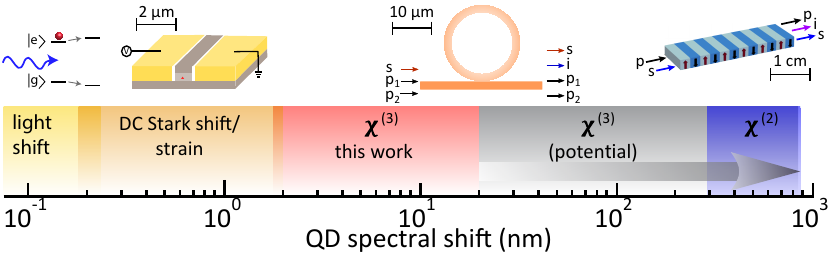}
		\caption{\textbf{Frequency shift techniques for quantum dots (QDs).} Relatively small shifts are typically achieved by tuning the QD energy levels, through optical fields (i.e., the light shift/AC Stark shift), strain, and electrical fields (DC Stark shift), as depicted on the left side of the image. The depicted ranges are typical results, but some engineered systems have produced significantly larger shifts~\cite{bennett_giant_2010,Pagliano2014,yuan_uniaxial_2018,tumanov_static_2018}.  Several hundred nanometer shifts have been obtained using quantum frequency conversion of the emitted photons in cm-scale $\chi^{(2)}$ nonlinear waveguides (right). Here, we implement four-wave mixing Bragg scattering, a $\chi^{(3)}$ non-linear process, in compact and power-efficient microring resonators, producing frequency shifts in an intermediate regime (red region) sufficient to cover the inhomogeneous broadening of QDs. Moreover, large spectral shifts can also be obtained through this process (gray area), enabling spectral shifts spanning from intraband to interband conversion (gray arrow).}
		\label{fig1}
	\end{figure*}

To generate identical photons from multiple QDs, one needs to overcome this spectral mismatch, and many different approaches have been considered.  Strain~\cite{seidl_effect_2006}, optical Stark shifts~\cite{bose_large_2011}, and electrical Stark shifts~\cite{findeis_photocurrent_2001} have been used to tune QD emission and enable interference of photons from different QDs~\cite{Flagg2010,Patel2010,he_indistinguishable_2013}. Through suitable engineering of the epitaxial growth layers or the device geometry surrounding the QD, the typical sub-nanometer wavelength shifts achievable by these approaches can be significantly increased to the $\approx$~10~nanometer scale~\cite{bennett_giant_2010,Pagliano2014,yuan_uniaxial_2018,tumanov_static_2018} (Fig.~\ref{fig1}).  However, these approaches may not be compatible with arbitrary photonic geometries, limiting the design space available when using such structures to achieve desired performance (e.g., in terms of Purcell enhancement, photon indistinguishability, and efficient collection into a desired optical channel). In contrast, quantum frequency conversion (QFC)~\cite{Kumar1990,Raymer2012} acts on the emitted photons rather than the QD energy levels, so that it can be applied to any arbitrary QD single-photon source geometry. QFC can achieve large spectral shifts, with upconversion~\cite{Rakher2010} and downconversion~\cite{Zaske2012,DeGreve2012} between telecom and near-visible photons emitted from QDs demonstrated in cm-scale, $\chi^{(2)}$ nonlinear waveguides. Along with their relatively large size and power consumption, such single-stage $\chi^{(2)}$ approaches necessitate large spectral shifts, and as a result both QD sources needed to be converted to a target wavelength far outside of the original band~\cite{Ates2012,Weber2018}.

In contrast, here we use four-wave mixing Bragg scattering (FWM-BS)~\cite{mckinstrie_translation_2005} in compact, power-efficient nanophotonic resonators~\cite{Li2016} to perform intraband conversion suitable for spectrally shifting the photons over a range between 1.6 nm and 12.8 nm, an appreciable fraction of the QD ensemble inhomogeneous distribution. Furthermore, as the spectral translation range in FWM-BS is set by the difference in frequencies of two pump lasers, it can also produce large spectral shifts, including downconversion to the telecom band at the single-photon level~\cite{Li2016}. FWM-BS thus provides a unique opportunity to cover an extremely large spectral translation range, including the gap between approaches that tune the QD energy levels and $\chi^{(2)}$ techniques (Fig.~\ref{fig1}).

In Ref.~\cite{Li2016}, our focus was on establishing the device engineering to enable efficient, microresonator-based FWM-BS, and experiments were restricted to working with classical input signals created by attenuated, continuous wave laser light. Here, we demonstrate true quantum frequency conversion of single-photon states produced by a QD.  We study how the linewidth of the QD photons influences the achievable conversion efficiency, due to the finite bandwidth of our frequency converter, and the impact of frequency conversion on photon statistics.  We also show how to tailor our frequency converter to work with a wide range of input wavelengths, of importance for addressing the inhomogeneous broadening of QDs on the same sample and across different wafer growths. Our results show the promise of integrated nanophotonics technology for quantum frequency conversion applications, while also highlighting future directions for improving device performance with respect to the key metrics of conversion efficiency and added noise.

\begin{figure*}[t]
	\includegraphics[width=\linewidth]{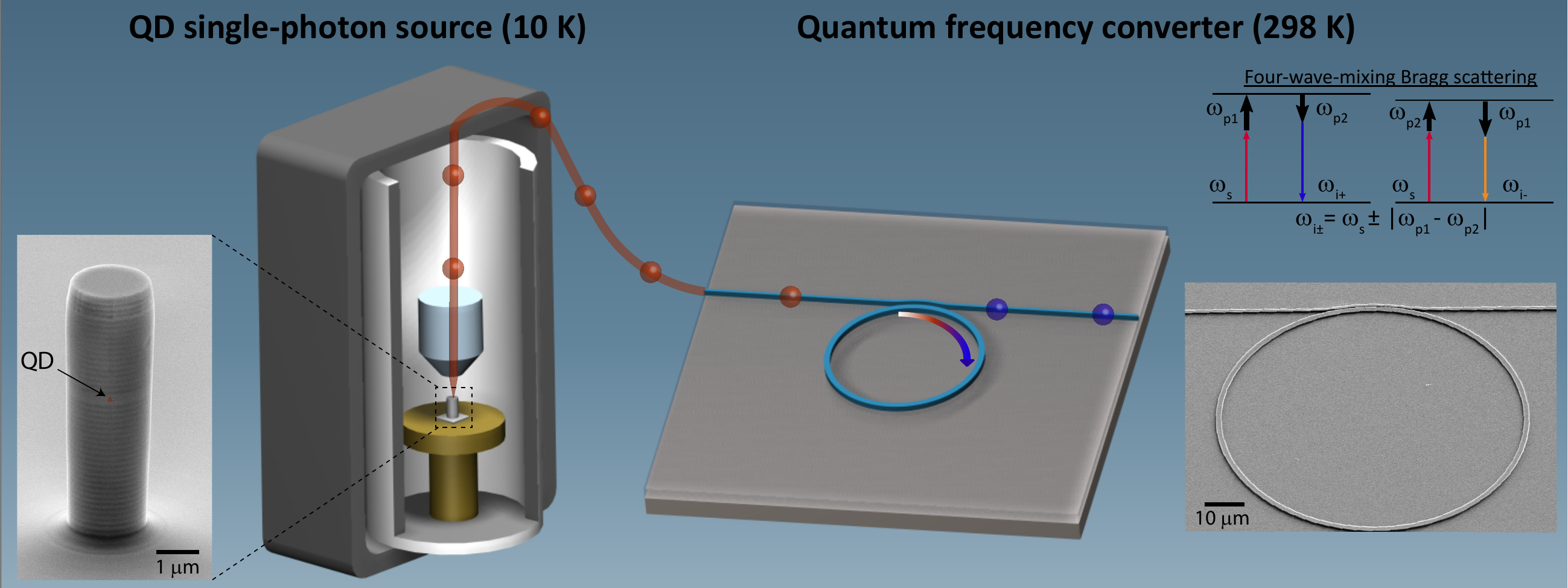}
	\caption{\textbf{Overview of the experiment.} Single photons from the source chip (QD in a micropillar cavity housed in a 10~K cryostat) are out-coupled via optical fiber and sent to a frequency converter chip (microring resonator) operating at room-temperature. An energy diagram depicting the four-wave mixing Bragg scattering process used for frequency conversion is shown in the top right, where two pumps ($\omega_{p1}$ and $\omega_{p2}$) shift the input signal ($\omega_{s}$) to idlers at frequencies $\omega_{i+}$ and $\omega_{i-}$. The output of the frequency converter is a superposition of the remnant (unconverted) signal and the two idlers, with filtering used to select a specific spectral channel. Scanning electron microscope images of the single-photon source and frequency converter are shown on the left and right sides of the image, with the inferred location of the QD indicated.}
	\label{fig2}
\end{figure*}

\section{QFC using FWM-BS}
To date, FWM-BS has been applied to quantum states of light produced by spontaneous nonlinear processes in macroscopic crystals and fibers, with QFC taking place within optical fibers~\cite{McGuinness2010,clemmen_ramsey_2016}.  Here, we combine a nanophotonic quantum light source - a single InAs/GaAs QD in a micropillar cavity - with a nanoscale frequency converter based on efficient and low-noise FWM-BS in Si$_3$N$_4$ micorings~\cite{Li2016}, as schematically depicted in Fig.~\ref{fig2}.  This first demonstration of QFC of QD single photons via FWM-BS highlights the optical compatibility of the source and frequency converter. This is non-trivial, as the frequency converter bandwidth must accommodate the source linewidth, while the temporal duration of the pumps that enable efficient frequency conversion must be longer than that of the single photon wavepackets. As described below, our microresonator-based frequency converter has a bandwidth on par with (and in some cases, significantly larger than) that of the photons generated by InAs/GaAs QDs~\cite{Aharonovich2016,Senellart2017}, and utilizes continuous-wave pumps (in contrast to ps and ns pulses used in Refs.~\cite{McGuinness2010} and~\cite{clemmen_ramsey_2016}, respectively), suggesting that these compatibility requirements can be met.  Moreover, the recent demonstration of heterogeneous integration of InAs/GaAs QD single-photon sources with Si$_3$N$_4$ nanophotonic circuits~\cite{Davanco2017} suggests that source and converter can eventually be combined within a single integrated chip.

The two-pump nature of FWM-BS means that input signal photons at $\omega_{s}$ can be both frequency up-shifted and down-shifted, with the shift given by the difference in pump frequencies $(\omega_{p1} - \omega_{p2})$, as shown in the energy diagram in Fig.~\ref{fig2}. The up- and down-shifted fields are referred to as the blue-shifted ($\omega_{i+}$) and red-shifted idlers ($\omega_{i-}$), respectively. The conversion efficiency into each idler depends on the degree to which the four fields involved are frequency-matched and phase-matched~\cite{Li2016}.  For our devices, both blue- and re-detuned idlers are generated with nearly equal efficiency, with a single idler selected by bandpass filtering of the output light.

\section{System Design}
Our microrings are fabricated in Si$_3$N$_4$, a material with low linear loss, appreciable nonlinearity, and negligible two-photon absorption at telecom wavelengths~\cite{Moss2013}. Our FWM-BS process involves two pumps in the telecommunications C-band (1525 nm to 1565 nm) that convert an input signal at $\approx$~917~nm to an output idler spectrally shifted between $\approx$~1.6 nm and $\approx$~12.8 nm from the input.  The resonator cross-section is chosen to ensure that the FWM-BS process is both phase-matched and frequency-matched; this is done by iterating between simulations that take into account material dispersion, waveguiding, and bending effects and experimental measurements of the cavity resonance positions using a wavemeter with specified 0.1~pm accuracy.  We engineer the parameters of the coupling waveguide (waveguide width, gap with respect to the ring, and interaction length) to achieve overcoupling at the signal and idler wavelengths, thus ensuring that the majority of input signal photons are coupled into the resonator, and the majority of frequency-converted idler photons are coupled back into the access waveguide. See supplementary secs.~VI and VII for more details.

\section{QFC of a QD Single-Photon Source on a Nanophotonic Chip}
\begin{figure}[!t]
	\includegraphics[width=\linewidth]{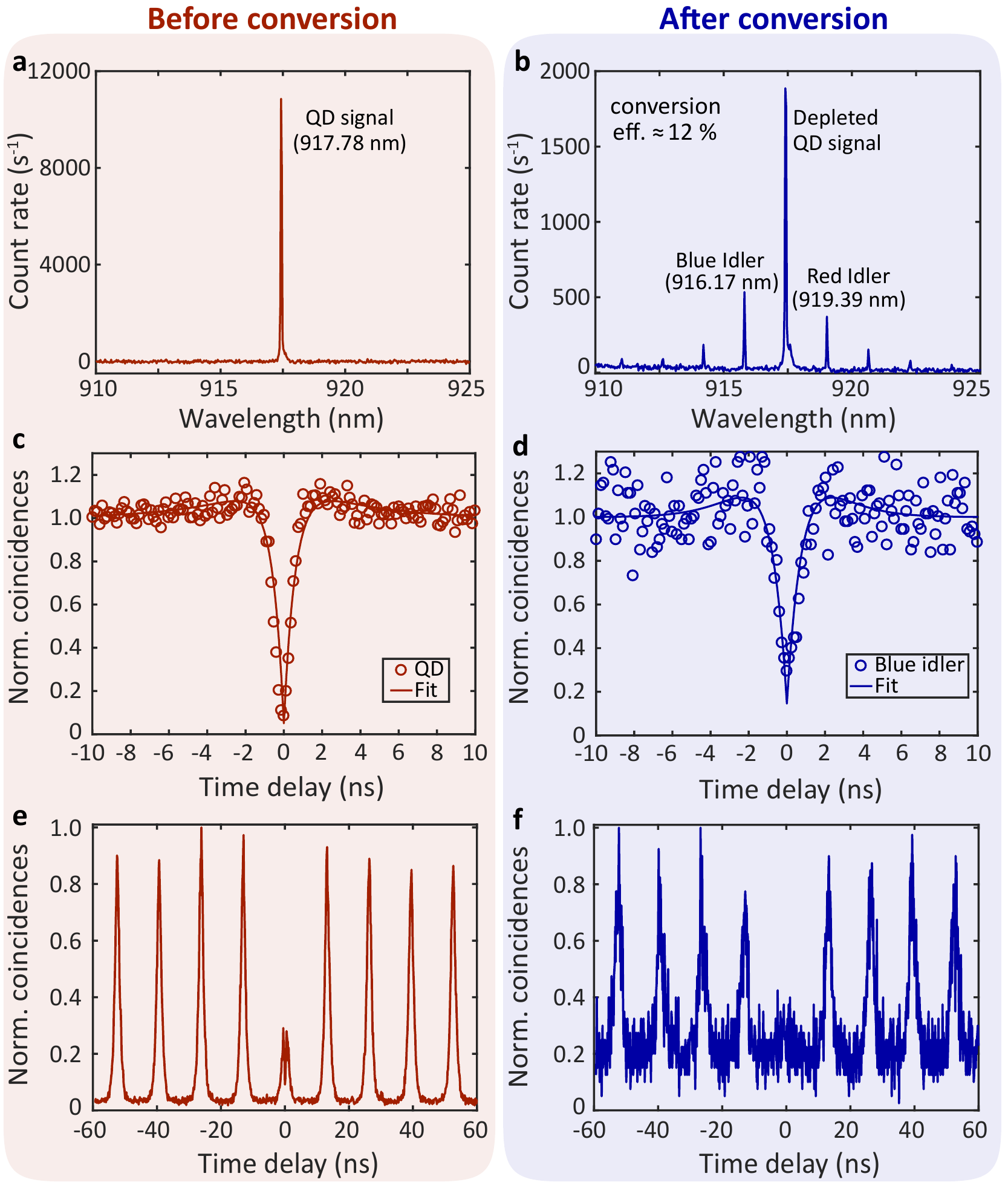}
	\caption{\textbf{Quantum frequency conversion of a QD single-phton source.} The left/right columns show measurement results before/after conversion, respectively. (a)-(b) Optical spectra for the two cases. The QD signal at 917.78 nm in (a) is sent to the frequency converter chip, whose output in (b) consists of the depleted signal and two dominant frequency-shifted idlers (blue idler at 916.17 nm and red idler at 919.39 nm). (c)-(d) The intensity autocorrelation of the QD is antibunched ($g^{(2)}(0)<0.5$) both before and after frequency conversion. Circles are data points and solid line is a fit to the data. (e)-(f) Intensity autocorrelation under pulsed excitation.}
	\label{fig3}
\end{figure}

We first spectrally shift our QD source using the microring frequency converter. The QD is excited at its p-shell at $\approx$~903.31 nm using a tunable continuous-wave laser (see supplementary sec.~I for info on the experimental setup and sec.~II for the QD source fabrication). The QD spectrum has a single peak at $\approx$~917.78 nm (Fig.~\ref{fig3}(a)), and so we temperature tune the frequency converter to match this wavelength, as discussed later in the context of Fig.~\ref{fig5}(c).  The QD emission is combined with two 1550 nm band pumps and sent into the microring converter, and the output spectrum of the converter shows a depleted QD signal that is accompanied by two dominant idlers, a blue idler at $\approx$~916.17 nm and red idler at $\approx$~919.39 nm (Fig.~\ref{fig3}(b)). The separation between either idler and the depleted QD signal is $\approx$1.61~nm ($\approx$~573.2~GHz), and is equal to the frequency difference between the two 1550~nm pumps, which was set to one free spectral range (FSR) of the microring resonator. The on-chip conversion efficiency is defined as the ratio of the frequency converted photon flux at the converter chip waveguide output to the input signal photon flux at the converter chip waveguide input. We estimate the conversion efficiency for the blue idler based on the two spectra (Fig.~\ref{fig3}(a),(b)) to be $12.8~\%~\pm~1.8~\%$, while photon counting (performed by measuring the photon flux in the input signal band and converted idler band) gives a conversion efficiency value of $11.4~\%~\pm~1.6~\%$, where the uncertainties are one standard deviation values due to fluctuations in the detected power and spread in the transmission of optical components in the experimental setup (see supplementary sec.~III for further discussion). The slightly higher conversion efficiency for the blue idler compared to the red idler is due its slightly better frequency matching, while the weak higher-order idlers in the spectrum are due to a pump mixing and cascaded frequency conversion effects~\cite{Li2016}.

The single-photon nature of the QD signal before and after frequency conversion is determined through measurement of its intensity autocorrelation $g^{(2)}(\tau)$ with a standard Hanbury-Brown and Twiss setup, where $\tau$ is the time delay between detection events on the two detectors. Under continuous wave (cw) excitation, the QD emits high-purity single photons with $g^{(2)}(0)=0.080\pm0.003$ (Fig.~\ref{fig3}(c)). The one standard deviation uncertainty in $g^{(2)}(0)$ is due to the fluctuation in the count rate on the detectors (supplementary sec.~IV). After frequency conversion, the blue idler remains antibunched with $g^{(2)}(0)=0.290\pm0.032$, see Fig.~\ref{fig3}(d). Thus, the light remains dominantly composed of single photons (relative to multiple photons) [i.e., g$^{(2)}(0) < 0.5$] after frequency conversion. The degradation of the antibunching dip is attributed to resonant noise generated by the 1550 nm pumps, potentially due to Si$_3$N$_4$ fluorescence. As discussed later, we operate the frequency converter in a high pump power regime to accommodate the relatively large QD linewidth, which comes at the expense of increased noise \cite{Li2016}.

The QD becomes a triggered single photon source when excited by a pulsed laser. The intensity autocorrelation of the QD under pulsed excitation (pulse width = 5 ps and the QD lifetime = 1 ns) is shown in Fig.~\ref{fig3}(e). Instead of a complete suppression of coincidences near zero time delay, the correlation curve has a small peak with a dip at zero delay, whose value is used in estimating $g^{(2)}(0)=0.10\pm0.06$. This behavior can likely be attributed to carrier recapture and multiple excitation of the QD within a pump pulse~\cite{Aichele2004,Singh2018}. The autocorrelation of the frequency-converted blue idler for pulsed excitation remains antibunched with $g^{(2)}(0)=0.31\pm0.07$ (Fig.~\ref{fig3}(f)). Similar to the cw case, noise from the converter chip results in an increase in $g^{(2)}(0)$, with the level of degradation similar in the two cases.  The pulsed measurement enables a clear attribution of the degradation in $g^{(2)}(0)$ to the frequency converter, as its noise is time-invariant and therefore not correlated with the QD itself, but is instead due to the cw pumps. The impact of converter noise, whose on-chip flux is estimated to be 1.5~$\times$~10$^4$~s$^{-1}$ and uniformly distributed in time, can be reduced if the on-chip QD photon flux is increased (e.g., through better coupling efficiency) or if we operate at lower pump powers, which is possible for a narrower linewidth QD source and will reduce the noise level. Further discussion of the frequency converter noise is in supplementary sec. VIII.

\begin{figure}[!t]
	\begin{center}
		\includegraphics[width=\linewidth]{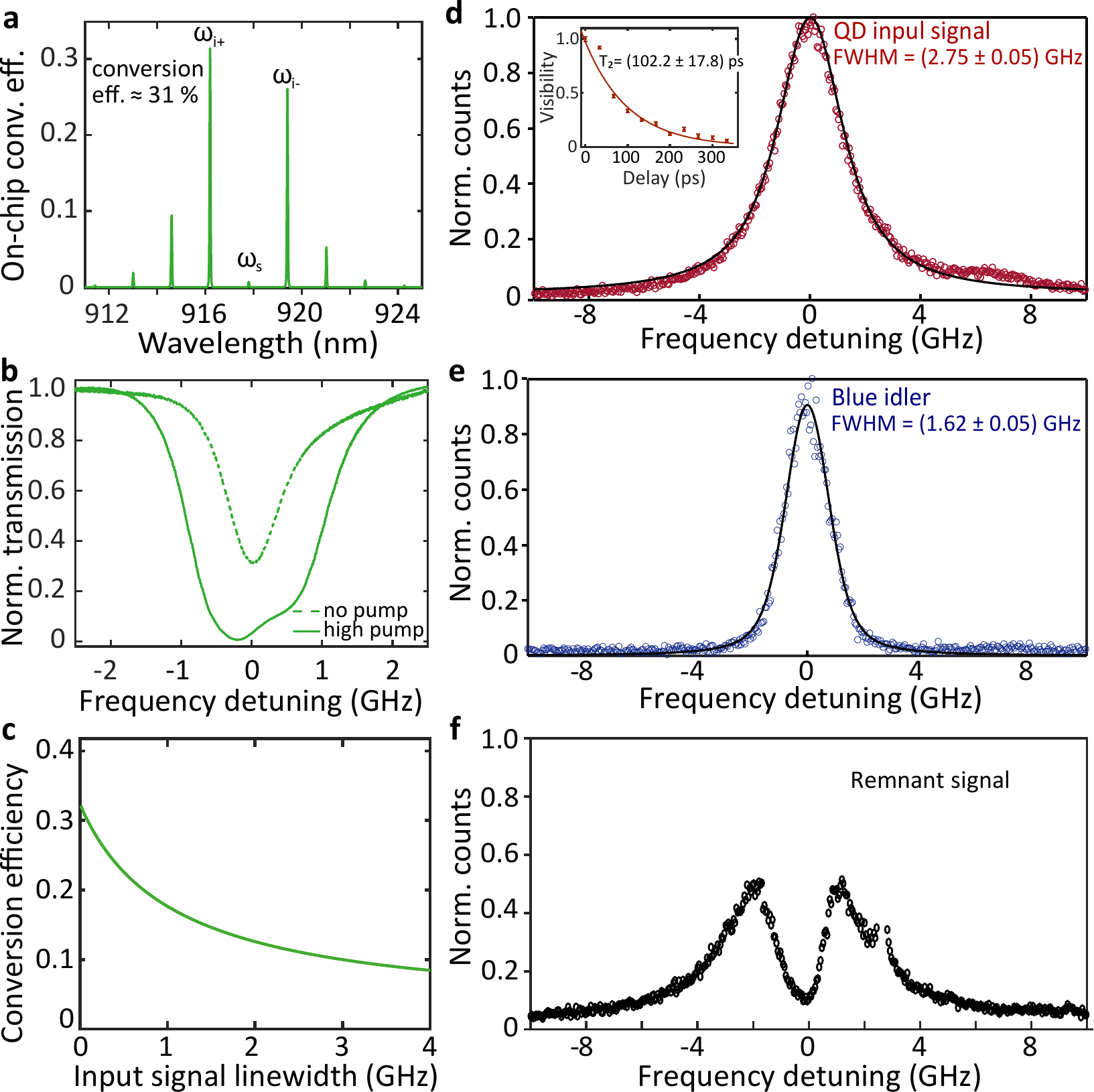}
		\caption{\textbf{Influence of QD linewidth on frequency converter performance.} (a) Frequency converter output spectrum for a narrow linewidth cw laser input. (b) Transmission spectrum of the microring frequency converter in the linear regime (with no pumps) and the nonlinear regime (total on-chip pump power $\approx$~20~mW) when scanned by a laser centered at 917 nm, providing an indication of the converter bandwidth. (c) Calculation of the expected conversion efficiency (green curve) as a function of input signal linewidth at a fixed linear linewidth for the microring frequency converter (1.12 GHz) and 1550~nm pump power (20 mW on-chip). (d) Measurement of the QD linewidth before frequency conversion, using a scanning Fabry-Perot resonator.  (inset) Measurement of the QD coherence time before frequency conversion, using an unbalanced Mach-Zehnder interferometer, normalized to the visibility at zero delay. The two measurements agree to within their uncertainties, which are one standard deviation values determined from nonlinear least squares fits to functional forms for the spectrum (Voigt) and coherence time (single-sided exponential). (e) Measurement of the frequency-converted blue idler linewidth, which is reduced relative to the linewidth in (d) due to the narrower frequency converter bandwidth. (f) The remnant QD signal (i.e., unconverted light) shows a dip in its spectrum as a result of the frequency conversion process. In (d)-(f) the circles are data points and the solid lines are Lorentzian fits to the data.}
		\label{fig4}
	\end{center}
\end{figure}

To determine the maximum attainable conversion efficiency if a narrower linewidth source is available, we substitute the QD source with a $\approx$~200~kHz linewidth cw laser, and measure the output of the frequency conversion chip on an optical spectrum analyser. Figure~\ref{fig4}(a) shows the measured spectrum, where the two prominent sidebands are the blue and red idlers ($\omega_{i+}$ and $\omega_{i-}$). In contrast to Fig.~\ref{fig3}(b), we observed that the conversion efficiency is significantly higher ($\approx$~31~$\%$ vs. $\approx$~12~$\%$ for the blue idler), and the signal ($\omega_{s}$) has been much more strongly depleted.  This suggests that the linewidth of the QD source is the cause of decreased conversion efficiency.  In sec. IX of the supplementary information, we discuss factors that limit the conversion efficiency to 31~$\%$, namely, conversion into multiple idlers rather than a single idler and non-unity out-coupling of converted light into the access waveguide. Correcting for these non-idealities should enable conversion efficiency approaching 90~$\%$.

We consider the role of source linewidth on conversion efficiency by scanning the narrow linewidth input laser across the cavity mode at 917~nm (Fig.~\ref{fig4}(b)).  The linear transmission spectrum (i.e., without application of the pumps) shows a linewidth of $\approx1$~GHz, which increases to $\approx2$~GHz at relatively high pump powers (10~mW per pump).  This suggests that the input source linewidth should be significantly narrower than 2~GHz to achieve full conversion efficiency (this is possible for QDs, for which the radiative-limited linewidth is $\approx$160~MHz for a lifetime of 1~ns). This prescription can be quantified by solving the coupled mode equations for the frequency converter (supplementary sec.~VII) with knowledge of a few experimentally-measured quantities (pump powers and microring intrinsic and coupling quality factors).  Figure~\ref{fig4}(c) shows the calculated conversion efficiency as a function of the input signal linewidth (green curve), assuming a Lorentzian frequency spectrum, and a loaded linear cavity linewidth of 1.12~GHz.  We see that the conversion efficiency is reduced by about a factor of three when going from a narrow-band input to a linewidth of 3~GHz.

With this guidance from theory, we next measure the QD source linewidth before and after frequency conversion, using a scanning Fabry-Perot (SFP) analyzer with a 200~MHz linewidth. Fitting to a Voigt profile, we measure a QD linewidth of $\approx$~2.75 GHz before frequency conversion (Fig.~\ref{fig4}(d)), and a coherence time of $\approx$~102~ps (see inset of Fig.~\ref{fig4}(d)) using an unbalanced Mach-Zehnder interferometer; the two values are consistent to within our measurement uncertainties. The two measurements are both subjected to the limitation of timing resolution that is slower than the typical spectral diffusion timescales for InAs/GaAs quantum dots~\cite{kuhlmann_charge_2013}, so that the measured linewidth/coherence time contain the influence of both dephasing and spectral diffusion processes~\cite{Kammerer2002,Coolen2007}. Referring back to the simulated conversion efficiency in Fig.~\ref{fig4}(c), for this input signal linewidth a conversion efficiency slightly more than 10~$\%$ is expected, close to the experimentally observed efficiency in Fig.~\ref{fig2}(b). Moreover, because the frequency converter has a narrower linewidth than the input QD photons, the frequency-converted light has a narrower linewidth, of $\approx 1.62$ GHz (Fig.~\ref{fig4}(e)).  As intuitively expected, the remnant QD signal (i.e., unconverted light), shows a dip in its spectrum (Fig.~\ref{fig4}(f)), further indicating the spectral filtering effect of the microresonator frequency converter.

\section{Versatility of the Frequency Converter}

\begin{figure*}[t]
	\includegraphics[width=\linewidth]{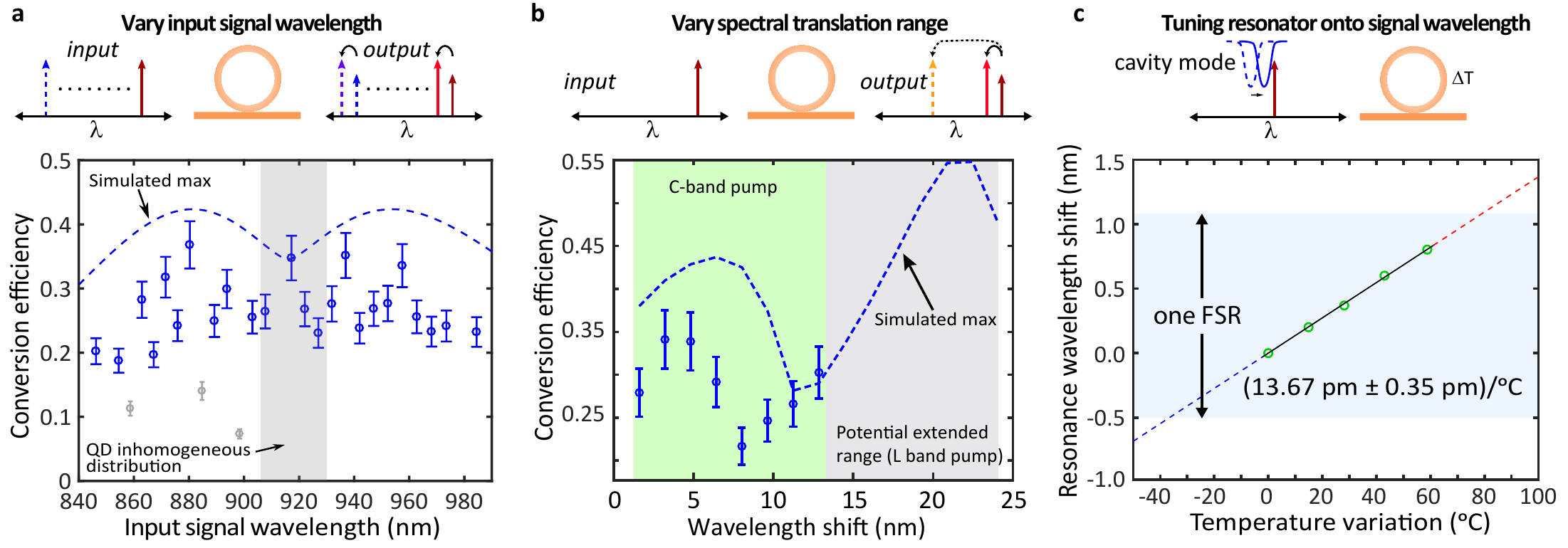}
	\caption{\textbf{Versatility of the frequency converter for QD applications.} (a) Conversion efficiency (open circles) as the input signal wavelength is varied while keeping the pump separation fixed. Gray data points correspond to wavelengths for which the microring frequency converter exhibits significant frequency mismatch due to mode interaction effects. The dashed curve is the simulated conversion efficiency, assuming experimentally estimated dispersion parameters and the assumption of fixed cavity quality factors. Error bars are one standard deviation uncertainties due to variations in fiber-to-chip coupling efficiency. (b) Conversion efficiency (open circles) as the frequency shift is varied by changing the spectral separation between the two pumps (input signal is fixed). The demonstrated range is limited to 12.8~nm (green shaded area), while a different choice of second pump laser is predicted to increase the range to $>$22~nm. Circles are data points and dashed line is a simulated curve. (c) Fine tuning of the nearest microring mode onto resonance with a fixed input signal through temperature. Circles are measured data points while dashed lines represent an extended temperature range (uncertainties in the measured data are smaller than the symbol size). A linear fit to the data gives a tuning rate of (13.67~pm~$\pm~$0.35~pm)/$^{\circ}$C, where the uncertainty is a 95~$\%$ confidence interval from the fit.}
	\label{fig5}
\end{figure*}

We further consider how our frequency converter can be used in making identical photons from multiple QDs.  As the input wavelengths can lie anywhere within the inhomogeneously-broadened QD distribution (typically a 10~nm to 50~nm spectral window), our converter must have a broad operating wavelength range. Figure~\ref{fig5}(a) shows this to be the case, with the conversion efficiency remaining $>$~20~$\%$ (mean value of 25~$\%$) over an exceedingly broad range of wavelengths from 840~nm to 980~nm (see supplementary Sec.~VI).  Here, we use a tunable, narrow linewidth cw laser as the input to determine the conversion efficiency in the limit of a narrow source linewidth, and the 1550~nm pumps are fixed at a one FSR separation. Measurements are compared against simulations (dashed line in Fig.~\ref{fig5}(a)), which account for the measured dispersion parameters of the microring converter, but assume fixed values of the microring intrinsic and coupling quality factors, equal to those measured for the 917~nm mode.  This is not true in practice, as the resonator-waveguide coupling and intrinsic quality factor varies with wavelength, and this is the main source of discrepancy between theory and experiment

We next consider the achievable spectral translation range. Use of a resonator means that, for a given device, frequency shifts are limited to integer multiples of the resonator FSR, modulo the resonator linewidth.  In Fig.~\ref{fig5}(b), we assess how the conversion efficiency changes as we vary this integer multiplier, which we measure by keeping the input signal and one pump near 1525~nm fixed, and varying the wavelength of the second pump in the C-band. Frequency shifts up to 4.5~THz (12.8 nm, corresponding to 8~FSRs) are achieved, with conversion efficiency between $\approx$~21~\% to $\approx$~34~\%. Measurements again differs from prediction (dashed line in Fig.~\ref{fig5}(b)) due to the assumption of fixed intrinsic and coupling quality factors for resonator modes.  We note that the spectral translation range is not limited by the device, but instead by the lasers available.  Using an L-band laser with coverage up to 1600~nm for the second pump results in predicted spectral shifts in excess of 8~THz (22.4 nm). In addition, the predicted conversion efficiency has significantly gone up.  This is due to an increased asymmetry in the degree to which both the blue and red idlers are nearly equally well frequency-matched.  A strong mismatch for one idler yields a higher conversion efficiency for the better-matched idler, as has been observed in practice in the case of wideband conversion in Ref.~\cite{Li2016}.

The frequency converter's discrete spectral resonances also require precise spectral matching between the input wavelength and an appropriate microresonator mode. As we have already shown that conversion efficiency is high for modes that span a broad range of input wavelengths (Fig.~\ref{fig5}(a)), we simply need to tune the nearest resonator mode to match the input signal. This is done by temperature tuning the microresonator, with any resulting spectral mismatch that occurs in the 1550~nm pump band compensated by tuning the individual pumps while keeping the pump separation fixed, so that the temperature tuning does not influence the spectral translation range. Figure~\ref{fig5}(c) shows tuning of the microring mode at 917~nm for a temperature change of up to $\approx$~60~$^{\circ}$C, over which an approximately linear shift of 13.7~pm/$^{\circ}$C is observed.  The total wavelength shift of $\approx$~820~pm is a bit more than half of the resonator FSR (1.6~nm).  Full FSR tuning would ensure that any input signal and a resonator mode could be matched, and can be achieved through further increase in the temperature, or by cooling, as the thermo-optic coefficient of Si$_3$N$_4$ remains nearly constant down to 200~K~\cite{elshaari_thermo-optic_2016}. Finally, the resonator FSR can be decreased to reduce the fundamental increment of frequency shift.  For our devices, the ring radius (which largely controls the FSR) can be increased without influencing phase/frequency matching (which is largely determined by the ring width and thickness).  One can thus envision many different frequency converter rings on the same chip, or even the same bus waveguide, where the rings differ in radius only, to provide different spectral translation increments.

\section{Conclusion}
In conclusion, we have demonstrated quantum frequency conversion of single photons from a QD using a nanophotonic frequency converter, with an on-chip conversion efficiency ($\approx$~12~$\%$) primarily limited by the linewidth of the QD source relative to the frequency converter bandwidth. Improved conversion efficiency can be obtained by using QD sources with sufficiently narrow linewidths (ideally a few times smaller than the converter bandwidth), or by increasing the loaded linewidth of the converter (supplementary sec.~VII). Future directions include demonstration of telecom-band downconversion and heterogeneous integration of the two elements using the approach developed in Ref.~\cite{Davanco2017}.  The ability to achieve $>$90~$\%$ transfer efficiency of single-photons from an InAs/GaAs QD single-photon source and a thick Si$_3$N$_4$ waveguiding layer shown in that work, together with the optical compatibility of QD single-photon source and Si$_3$N$_4$ microring frequency converter shown in this work suggests a future route to single-chip integration, a critical step when identical photons from multiple QDs on the same chip are needed.

\section*{Acknowledgments}
A. Singh, Q. Li, X. Lu acknowledge support under the Cooperative Research Agreement between the University of Maryland and NIST-PML, Award 70NANB10H193. C. Schneider and S. H{\"o}fling acknowledge support by the State of Bavaria and the BMBF within the project Q.Com-HL. C. Schneider acknowledges funding by the DFG (Project SCHN1376 5.1). S. Liu, Y. Yu, and J. Liu acknowledge the National Key R\&D Program of China (2018YFA0306100), the National Natural Science Foundations of China (11874437,11704424), the Natural Science Foundation of Guang-dong Province (2018B030311027,2017A030310004,2016A030310216), and the Guangzhou Science and Technology project (201805010004).


\onecolumngrid \bigskip
\appendix
\setcounter{figure}{0}
\setcounter{equation}{0}
\makeatletter
\renewcommand{\theequation}{S\@arabic\c@equation}
\begin{center} {{\bf \large SUPPORTING
INFORMATION}}\end{center}

\renewcommand{\thefigure}{S\arabic{figure}}
\renewcommand{\thesection}{\Roman{section}}

\section{Experimental setup}

\begin{figure*}[t]
	\centering
	\includegraphics[width=14cm]{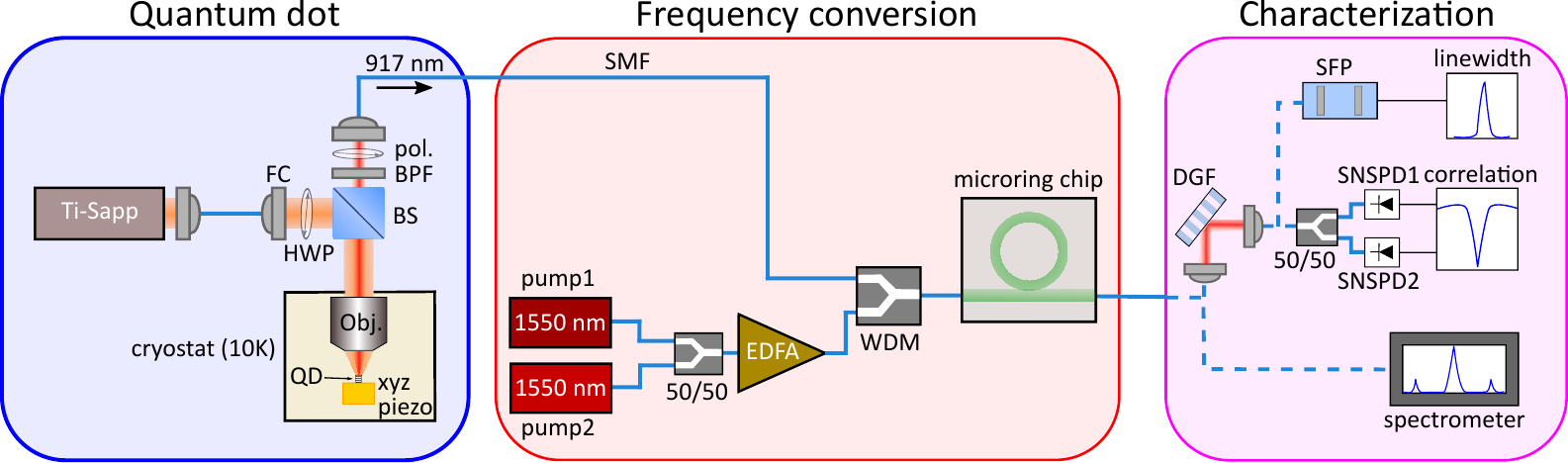}
	\caption{The experimental setup consists of the QD source in a closed-cycle cryostat at 10 K and the frequency converter chip is in a separate room-temperature setup. The QD single photons before and after frequency conversion are characterized through a grating spectrometer, intensity autocorrelation using a time-correlated single photon counting (TCSPC) module, and scanning Fabry-Perot cavity for the linewidth measurement. FC: Fiber coupler, BPF: Band-pass filter, HWP: half-wave plate, BS: Beam splitter, pol: polarizer, SMF: Single-mode fiber, EDFA: Erbium-doped fiber amplifier, WDM: Wavelength division multiplexer, DGF: Diffraction grating filter, SNSPD: Superconducting nanowire single photon detector, SFP: Scanning Fabry-Perot.}
	\label{figS1}
\end{figure*}

Figure~\ref{figS1} shows a sketch of the experimental setup used for the quantum dot (QD) quantum frequency conversion (QFC) experiment. The QD-micropillar source is placed inside a closed-cycle cryostat operating at 10~K, where the QD is excited in its p-shell at 903.31 nm (measured on a wavemeter) by a tunable Ti-Sapphire cw laser. The photons are collected through a high numerical aperture objective (NA = 0.75, 100x) that sits within the cryostat, enabling high efficiency collection of emitted photons \cite{Liu2017}. A half-wave plate (HWP) enables selection of the laser polarization incident on the sample, and an output polarizer (pol.) in the detection path can select for a specific emission polarization and help suppress laser scatter. A narrow (1~nm) band-pass filter separates the excitation from the emission, which is coupled to a single-mode fiber (SMF) and sent to the frequency converter chip in a separate room-temperature setup. The frequency converter chip is temperature-tuned to match its nearest cavity mode with the QD emission. The two 1550 nm pumps are combined in a 50/50 coupler and sent to an erbium-doped fiber amplifier (EDFA) to generate required total pump power ($\approx$ 20~mW on the chip). The QD signal and amplified pumps are combined by a wavelength division multiplexer (WDM), and injected into the microring access waveguide through a lensed fiber. The frequency-converted signals are collected through another lensed fiber at the access waveguide output. The converted signal can be directly coupled to a grating spectrometer for spectral analysis, or into a ruled reflective diffraction grating optical setup of bandwidth $\approx$ 150 GHz (efficiency $\approx$~50 \%) to spectrally select the blue-shifted idler from the frequency-converted spectrum. We use a standard Hanbury-Brown and Twiss setup to measure the intensity autocorrelation of the QD emission before and after frequency conversion, where photons are detected using superconducting nanowire single photon detectors (SNSPDs) operating in a 0.7~K cryostat. The linewidth of the signal and idler is measured using a separate setup based on a scanning Fabry-Perot cavity with free-spectral range of $\approx$~40 GHz and linewidth of $\approx$~200 MHz.

\section{Fabrication of the Micropillar Quantum Dot Source}
The QD sample consists a single layer of low density InAs QDs grown via molecular beam epitaxy and located at the center of a $\lambda$-thick GaAs cavity surrounded by two Al$_{0.9}$Ga$_{0.1}$As/ GaAs Bragg mirrors with 12 (25) pairs. The density of self-assembled InAs quantum dots varies continuously along the wafer by stopping the rotation of the substrate during InAs deposition.

The first step in fabrication of the micropillar cavities is location of the QDs using a photoluminescence-based positioning technique~\cite{Liu2017a}. Next, the sample is spin-coated with a negative tone electron beam resist (HSQ FOx15). The resist is exposed using an electron-beam lithography system at 100 keV. After the exposure and development processes, the mask pattern is transferred into the sample via an inductively-coupled plasma reactive ion etching system.

\section{Estimation of conversion efficiency}

The conversion efficiency of the blue idler is estimated from the experimental measurements in two ways, using spectrometer data and through photon counting. In case of the spectrometer measurements, we integrate the area under the peak of the QD input signal and the frequency-converted blue idler (see Fig.~3a,b of main text) to obtain the total counts before and after conversion. The QD input signal spectrum (Fig.~3a) is obtained after spectrally filtering the excitation laser through a grating filter (transmission efficiency of $\approx$~50~\%, Figure~\ref{figS1}), whereas the grating filter is bypassed during the conversion measurement (i.e., the QD signal is sent directly to the converter chip). The spectrum after conversion is obtained by directly sending the converted signal to the spectrometer without any spectral filtering (Fig.~3b). Accounting for the differences in filtering and the chip transmission efficiency of $\approx$~15~\%, the on-chip conversion efficiency for the blue idler comes out to be $12.8~\%~\pm~1.8~\%$, where the uncertainty is a one standard deviation value that arises from the spread in chip coupling losses and transmission of the grating filter, as well as fluctuations in the measured spectrometer count rates.  For photon counting, we spectrally filter both the QD signal and blue idler before sending them to the SNSPDs. The photon counts are recorded during the intensity autocorrelation measurements. The conversion efficiency of the blue idler in this case is found to be $11.4~\%~\pm1.6~\%$, where the uncertainty is a one standard deviation value that arises from the spread in chip coupling losses and transmission of the grating filter, as well as fluctuations in the count rates measured by the SNSPDs.  The conversion efficiency value for the two cases are similar to within the measurement uncertainties, and match the theoretically predicted value as shown in Fig.~4c.

\section{Analysis of Intensity Autocorrelation data}
We use a standard Hanbury-Brown and Twiss setup to obtain the second-order correlation function $g^{(2)}(\tau)$ of the QD signal before and after frequency conversion, where $\tau$ is the time delay between detection events on the two SNSPD (Fig~S1). We record histograms of delays between detection events using a time-correlated single photon counting (TCSPC) module with 128~ps resolution. For sufficiently small $\tau$, the histogram of coincidences is equivalent to $g^{(2)}(\tau)$, see \cite{Lounis2000a}. The normalized histogram under cw excitation is fitted by a double-exponential function:

$$\textrm{g}^{(2)}(\tau) = 1 + A_1\text{exp}(-\gamma_1\cdot|\tau|) + A_2\text{exp}(-\gamma_2\cdot|\tau|)$$

\noindent with $A_1 + A_2 = -1$. This form is expected for a two-level system coupled to a single dark state, and describes antibunching at $\tau = 0$, bunching at some later time delay, and a return to the Poissonian level at $\tau \to \infty $ \cite{Davanco2014}. The fits shown in Fig.~3c,d of the main text are obtained using a nonlinear least squares procedure (for the pulsed pumping data, only the raw data is presented and no fit is performed). In all cases, the $g^{(2)}(0)$ value is taken as the measured data point at zero time delay and not the fit value. For the error in the $g^{(2)}(0)$ values, we calculate the fluctuation in the coincidence counts in the histograms for $\tau \gg 0$ (i.e., the Poissonian level), and propagate the error.

\section{Fabrication of the Microring Frequency Converter}
Microrings are fabricated in a 500~nm thick Si$_3$N$_4$ layer on top of a 3~$\mu$m thick SiO$_2$ layer. The Si$_3$N$_4$ layer was created by low-pressure chemical vapour deposition while the SiO$_2$ layer was grown via thermal oxidation of a 100~mm Si wafer. The wavelength-dependent refractive index and thickness of the layers were determined using a spectroscopic ellipsometer, with the data fit to an extended Sellmeier model. After cleaving into chips, the microring-waveguide devices were created by electron-beam lithography of a negative tone resist, followed by reactive ion etching of the Si$_3$N$_4$ using a CF$_4$/CHF$_3$ chemistry, removal of deposited polymer and remnant resist, and annealing at 1150~$^{\circ}$C in an O$_2$ environment for 3~h. The  microring used in our experiments has radius of 40~$\mu$m and a ring width of 1450~nm.

\section{Frequency converter design summary}

While four-wave-mixing Bragg scattering (FWM-BS) has been demonstrated in optical fibers in several experiments (e.g., ~\cite{McGuinness2010,clemmen_ramsey_2016}, it has only recently been demonstrated in integrated microresonators~\cite{Li2016}.  We qualitatively review the design approach for these devices here, and summarize some of the salient features with respect to the quantum dot quantum frequency conversion experiments that are the focus of this work.

FWM-BS uses two non-degenerate pumps at frequencies $\omega_{p1}$ and $\omega_{p2}$ to spectrally shift an input signal at $\omega_{s}$ to output idlers, one of which is blue-shifted with respect to the signal, that is, $\omega_{i+} = \omega_{s} + |\omega_{p1}-\omega_{p2}|$, and the other which is red-shifted with respect to the signal, $\omega_{i-} = \omega_{s} - |\omega_{p1}-\omega_{p2}|$.  Like any other $\chi^{(3)}$ parametric nonlinear optical process~\cite{agrawal_nonlinear_2007}, efficient operation means that these energy conservation relationships need to be accompanied by momentum conservation, or phase-matching, so that $\beta_{i+} = \beta_{s} + |\beta_{p1}-\beta_{p2}|$ and $\beta_{i-} = \beta_{s} - |\beta_{p1}-\beta_{p2}|$, where $\beta_{k}$ is the propagation constant of light at frequency $\omega_{k}$.  $\beta_{k} = 2\pi n_{\text{eff},k}/\lambda_{k}$, where $n_{\text{eff},k}$ is the effective refractive index at wavelength $\lambda_{k}$.  $n_{\text{eff},k}$ is a wavelength-dependent quantity influenced by different factors, including the material refractive indices of the waveguide layers and the effects of geometric confinement. In most cases in which FWM-BS has been studied in optical fibers~\cite{uesaka_wavelength_2002,mckinstrie_translation_2005,mechin_180-nm_2006,McGuinness2010,lefrancois_optimizing_2015}, the dispersion of the fibers was such that only one of the two possible idlers could be both frequency- and phase-matched, so that only a single idler needed to be considered. In cases in which the pumps are relatively close to each other spectrally, however, both up- and down-shifted idlers can satisfy these criteria~\cite{agha_low-noise_2012}, so that the conversion efficiency into each idler can be approximately equal.

We use the microring geometry to provide resonant enhancement of the FWM-BS process, enabling efficient conversion to be achieved for ten mW-scale, continuous wave pumps~\cite{Li2016}. The phase-matching criterion  $\beta_{i\pm} = \beta_{s} \pm |\beta_{p1}-\beta_{p2}|$ becomes $m_{i\pm} = m_{s} \pm |m_{p1}-m_{p2}|$, where $m_{k}$ is the azimuthal mode number of the cavity mode in the relevant band (signal, idler, pump 1, or pump 2).  While finding a set of whispering gallery modes that satisfies this relationship is straightforward, one then needs the corresponding mode frequencies to satisfy the energy conservation relationship $\omega_{i\pm} = \omega_{s} \pm |\omega_{p1}-\omega_{p2}|$, which will not be true for an arbitrary resonator cross-section.  As described in Ref.~\cite{Li2016}, we can satisfy this frequency matching requirement by looking for resonators that have equal free-spectral ranges in the two wavelength bands (pumps in the 1550~nm band and signal/idler in the 920~nm band).  For our devices, this prescription leads to relatively weak dispersion in the two wavelength bands. This enables a large number of mode combinations to be phase- and frequency-matched, providing significant flexibility in the operation of the frequency converter, as described in Fig.~5 in the main text.  In particular, this weak dispersion is the key that allows efficient conversion for both a wide range of signal wavelengths and a tunable spectral translation range (Fig.~5a,b in the main text).

\section{Modeling of the Microring Frequency Converter and Its Efficiency}
Modeling of the FWM-BS process in microrings was developed in Ref.~\cite{Li2016}; see also Ref.~\cite{Vernon_QFC}.  The most general approach that considers both FWM-BS and any competing four-wave-mixing processes is that based on numerical simulation of the Lugiato-Lefever equation, as discussed in ~\cite{Li2016}.  While such an approach accounts for the higher-order idlers we observe in experiment, the basic behavior of the system, for example the conversion efficiency into the first-order idlers, can be well-described by a simplified coupled mode theory that considers a restricted basis of modes.  Here, we first outline this theory, before describing how we apply it to photons emitted from single QDs, so that we can calculate the conversion efficiency as a function of input signal linewidth as shown in Fig.~4c of the main text. We also consider how, if the input signal linewidth is fixed, the loaded linewidth of the frequency conversion ring can be adjusted to improve conversion efficiency.

We start with the coupled mode equations \cite{Li2016}:

\begin{gather}
	t_R\frac{d E_{s}}{dt}=-(\alpha + i\delta_s )E_s + i\Omega_0 E_{i-} + i \Omega_0 E_{i+} + i \sqrt{\theta P_s}, \label{Eq_CMT_s} \\
	t_R\frac{d E_{i+}}{dt}=-\left(\alpha + i(\delta_s + \Omega_1+\Omega_2)\right)E_{i+} + i \Omega_0 E_s, \label{Eq_CMT_i1} \\
	t_R\frac{d E_{i-}}{dt}= -\left(\alpha + i (\delta_s - \Omega_1 +\Omega_2)\right)E_{i-} + i \Omega_0 E_s,
	\label{Eq_CMT_i2}
\end{gather}
where $E_{s,i\pm}$ are the intracavity mean fields corresponding to the signal and two adjacent idlers ($|E|^2$ representing the average power traveling inside the cavity), $t_R$ is the round-trip time, $\alpha$ is the cavity loss rate in the 930 nm band ($\alpha=\hat{\omega}_st_R/(2Q_L)$ with $\hat{\omega}_s$ and $Q_L$ being the signal resonance frequency and its loaded $Q$, respectively), $\delta_s$ denotes the signal detuning, $\theta$ is the power coupling coefficient between the resonator and the access waveguide ($\theta=\hat{\omega}_st_R/Q_c$ with $Q_c$ being the coupling $Q$), and $P_s$ represents the power of a cw signal. The parameters $\Omega_n (n=0,1,2)$ are defined as:
\begin{gather}
	\Omega_0 \equiv 2\gamma_{s}L|E_{p1}E_{p2}|, \label{Eq_Omega0}\\
	\Omega_1 \equiv \frac{\delta_{i+}-\delta_{i-}}{2},  \label{Eq_Omega1}\\
	\Omega_2  \equiv \frac{\delta_{i+}+\delta_{i-}-2\delta_s}{2}, \label{Eq_Omega2}
\end{gather}
where $\gamma_s$ is the Kerr nonlinear coefficient in the 930 nm band, $L$ is circumference of the microring resonator ($L\equiv 2\pi R$ with $R$ being the ring radius), $E_{p1,p2}$ denote the intracavity mean fields of the two pumps in the 1550 nm band, and $\delta_{i\pm}$ are the detunings of the two idlers. A straightforward calculation shows that $\Omega_{1,2}$ can be expressed as:
\begin{gather}
	\Omega_1  \approx  \left(D_1 |\mu| - |\omega_{p1}-\omega_{p2}|\right)t_R - \frac{\gamma_p L}{2} \left(|E_{p1}|^2 - |E_{p2}|^2\right),  \label{Eq_Omega1a}\\
	\Omega_2  \approx \frac{1}{2}D_2 \mu^2 t_R,\label{Eq_Omega2a}
\end{gather}
where $\gamma_p$ is the Kerr nonlinear coefficient in the 1550 nm band. In our configuration, we use two pumps with equal power, thus $\Omega_1$ (Eq.~\ref{Eq_Omega1a}) is determined by the difference in frequency of the pump lasers ($|\omega_{p1}-\omega_{p2}|$) and the FSR in the 930 nm band ($D_1$).

The above set of equations serves as a convenient tool for us to simulate the on-chip conversion efficiency for both the cw and pulsed input signal. First, for the cw case we can simply calculate the steady-state solution of the intracavity fields such as $E_{i\pm}$ for a given signal power and detuning. The corresponding power of the idlers in the waveguide is calculated based on input-output relations, which are subsequently normalized by the input signal power to give the on-chip conversion efficiency. Next, for the pulsed input, the signal is decomposed into a series of cw components through Fourier analysis. The spectrum of the converted idler can then be obtained by solving the steady-state idler fields for each cw component. The conversion efficiency in this case is defined as the averaged idler photon flux normalized by the averaged signal photon flux.

\begin{figure}[h]
	\includegraphics[width=8cm]{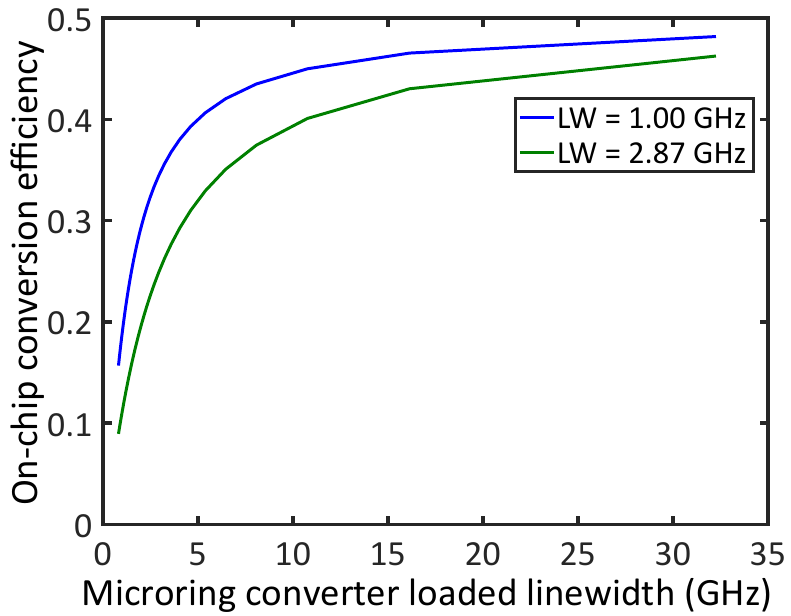}
	\caption{Simulated conversion efficiency for increasing microring frequency converter loaded linewidth for fixed input signal linewidth (LW) of 1 GHz (blue) and 2.87 GHz (green). The green curve represents the linewidth of the QD studied in the main text.}
	\label{figS2}
\end{figure}

We now consider the case in which the loaded linewidth of the microring frequency converter is a free parameter, with a fixed intrinsic quality factor ($Q$) of $1.6\times 10^6$. In practice, the loaded linewidth can be varied by tailoring the access waveguide coupling design.  We consider the conversion efficiency for input signals with two different linewidths, the 2.87~GHz linewidth we measure for our existing QD single-photon source, and a 1.0~GHz linewidth that has been achieved for many QD single-photon sources reported in the literature (the transform-limited linewidth for a system with a 1~ns lifetime is $\approx$~160~MHz). For a fixed input signal linewidth, the conversion efficiency increases with the converter bandwidth (Fig.~\ref{figS2}), at a relatively fast rate up until the point at which the microring resonator loaded linewidth is a factor of two or three times larger than that of the input signal, and then slowly thereafter until the conversion efficiency saturates.  For the QD studied in the main text, the conversion efficiency can increase from the 12~$\%$ we measure to $\approx$30~$\%$, if the microresonator loaded linewidth is increased by about a factor of four, to $\approx$4.5~GHz.  This will primarily come at the expense of the pump power required for efficient conversion, assuming that the loaded quality factors in the 1550~nm pump band are correspondingly reduced (we note that it may be possible to engineer the waveguide coupling so that only the input signal and output idler resonances have 4.5~GHz linewidths, while the 1550~nm pump resonances remain close to critically coupled at their narrower linewidths).  On the other hand, for the 1~GHz linewidth QD, a modest increase in loaded linewidth to 2~GHz would result in a 30~$\%$ conversion efficiency.

\section{Frequency converter noise}
The 1550~nm pumps used in our frequency converter induce added noise in the 917~nm band, resulting in degradation in the g$^{(2)}(0)$ values as observed in Fig.~3 in the main text. In this section, we further quantify this on-chip noise flux and its effect on the performance of QD single-photon sources.

For a total on-chip pump power of 20~mW, we measure an on-chip noise flux in the blue-shifted idler band of 3.2~fW~$\pm$~0.1~fW, where the one standard deviation uncertainty is due to variation in the chip insertion loss.  This corresponds to an on-chip noise photon flux of 1.5~$\times$~10$^4$ s$^{-1}$, which as mentioned in the main text, is uniformly distributed in time.  For scenarios in which the QD is operated under pulsed excitation, this can be converted to a number of noise photons per excitation pulse.  At an 80 MHz repetition rate and assuming a 2~ns time bin (consistent with containing the full wavepacket of a QD single photon), this corresponds to 3~$\times$~10$^{-5}$ noise photons per pulse.  A QD single-photon source which not only has a high source brightness but efficient coupling to the frequency converter chip might be expected to generate at least 0.01 photons per pulse (on-chip), resulting in a signal-to-noise level in excess of 30 (with a maximum value of 3~$\times$~10$^3$) at 10~$\%$ conversion efficiency.  Moreover, as noted in the main text, we operate at a relatively high pump power to broaden the converter bandwidth, due to the relatively broad linewidth photons our QD source generates.  At a pump power of 10~mW, the same conversion efficiency can be achieved if the source linewidth is sufficiently narrow, with the noise flux reduced by about a factor of 2.5.

\begin{figure}[h]
 	\includegraphics[width=8cm]{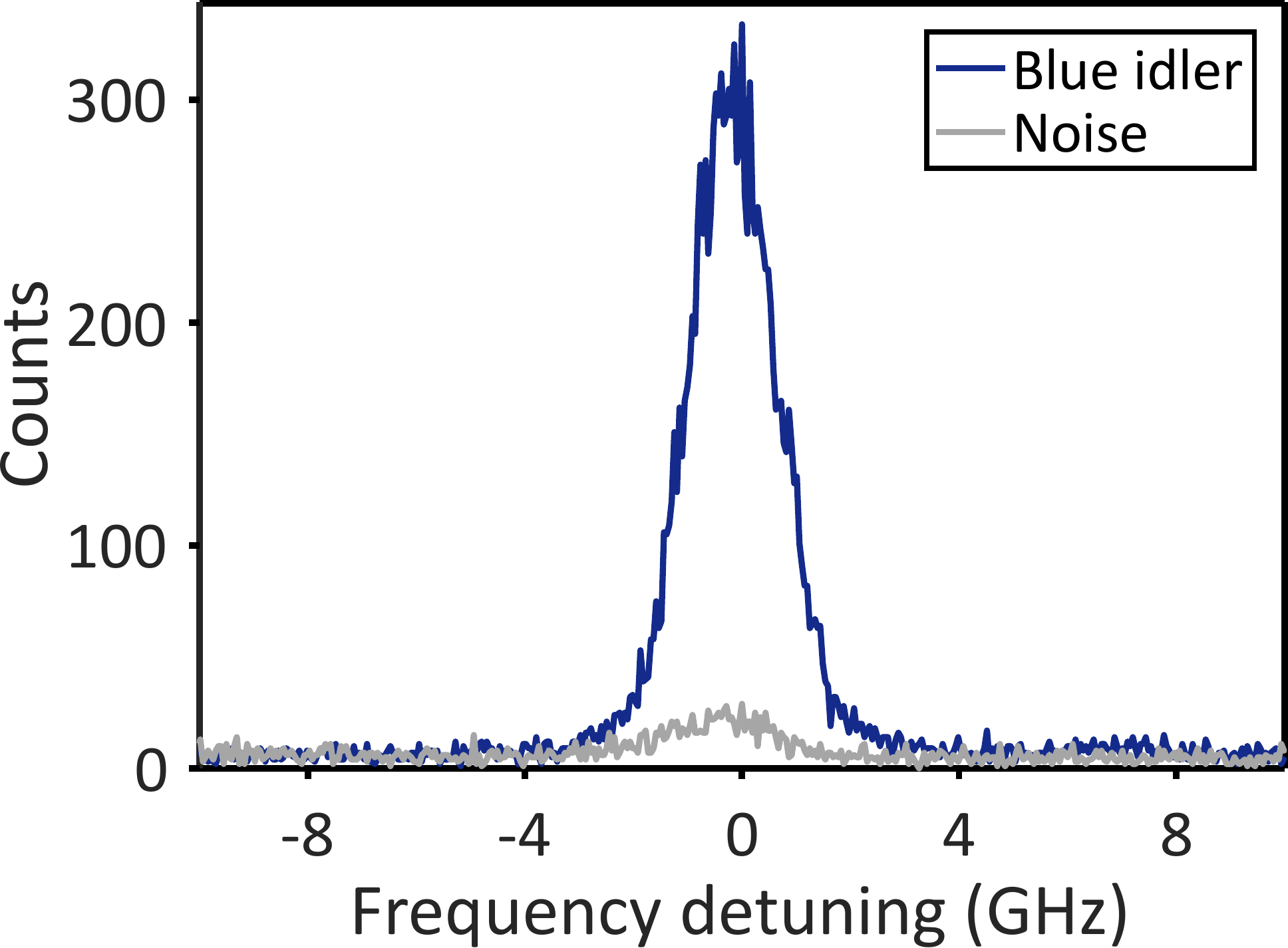}
	\caption{Scanning Fabry-Perot measurement of the converted blue idler (blue) and the converter noise (grey) in the same channel. The noise is measured in absence of the input QD signal.}
	\label{figS3}
\end{figure}

To provide some additional perspective on the added noise, we measure its spectrum in the converted idler channel using a scanning Fabry-Perot (SFP) resonator, as discussed in the main text. The QD signal is disconnected from the converter chip while the pumps remain on, and the added noise signal is measured (after the bandpass grating filter that initially selects this spectral channel), using the same conditions as the blue idler bandwidth measurement. The resulting spectrum of the added noise is plotted alongside the spectrum of the converted blue idler in Fig.~\ref{figS3}, and exhibits a signal-to-noise ratio of more than 10.  We note that if sufficiently narrow linewidth input photons sources are used (e.g., from a QD source exhibiting indistinguishable photons or other lifetime-limited quantum emitters such as vacancy centers in diamond or single alkali atoms), additional spectral filtering within the microcavity frequency converter bandwidth can provide improved noise performance.

\section{Optimizing the frequency converter efficiency}
Here, we discuss the factors affecting the conversion efficiency and its optimum limit. There are essentially three factors that are limiting conversion efficiency. First, both the blue-shifted and red-shifted idlers are created with roughly equal conversion efficiency, as they are both equally well phase- and frequency-matched (though this balance can be slightly adjusted by tuning the pump spectral positions). From Fig.~4a (main text), we can precisely add up the conversion efficiency for the two idlers in the experiment, which would be 31~\% (blue-shifted) + 26~\% (red-shifted) = 57~\%. Next, we have conversion into higher-order idlers. From Fig.~4a, conversion into higher order idlers adds up to about 16~\%. Therefore, the total conversion efficiency into all idlers (first order red- and blue-detuned plus all higher order idlers) is about 73~\%.

Finally, we need to take into account the level of waveguide-resonator coupling, that is, the requirement that (ideally) all input photons be coupled into the microring resonator, and all frequency-converted output photons be coupled out of the microring. Our microring's total loaded linewidth is $\approx$~1.1~GHz, of which $\approx$~200~MHz represents the intrinsic linewidth and the remaining 900~MHz is due to coupling to the access waveguide. On the extraction side, this corresponds to out-coupling about 82~\% of the frequency-converted photons from the ring into the access waveguide. The 73~\% conversion efficiency for all idlers mentioned above could thus be improved to 89~$\%$ if perfect overcoupling is achieved.

The above analysis suggests that a conversion efficiency approaching 90~$\%$ could be achieved provided that the level of overcoupling is improved, and that only one set of microring modes be phase- and frequency-matched.  In practice, this might be possible by using different techniques to mismatch undesired conversion channels (i.e., the red-shifted idler and higher-order idlers). Mode selective gratings and coupled resonators are amongst the approaches that might enable such possibilities.

\end{document}